\documentclass[aps,pra,multicol,epsf,epsfig,float,showpacs,showkeys]{revtex4}
\usepackage{graphicx}
\usepackage{dcolumn}
\usepackage{bm}

\begin{document}
\newcommand{\ri}{{\rm i}}
\newcommand{\re}{{\rm e}}
\newcommand{\bx}{{\bf x}}
\newcommand{\br}{{\bf r}}
\newcommand{\bk}{{\bf k}}
\newcommand{\bE}{{\bf E}}
\newcommand{\bR}{{\bf R}}
\newcommand{\bn}{{\bf n}}
\newcommand{\bd}{{\bf d}}
\newcommand{\rSi}{{\rm Si}}
\newcommand{\beps}{\mbox{\boldmath{$\epsilon$}}}
\newcommand{\rg}{{\rm g}}
\newcommand{\tr}{{\rm tr}}
\newcommand{\xmax}{x_{\rm max}}
\newcommand{\ra}{{\rm a}}
\newcommand{\rx}{{\rm x}}
\newcommand{\rs}{{\rm s}}
\newcommand{\rP}{{\rm P}}
\newcommand{\up}{\uparrow}
\newcommand{\down}{\downarrow}
\newcommand{\hc}{H_{\rm cond}}
\newcommand{\kb}{k_{\rm B}}


\title{Entanglement from thermal black body radiation} 
\author{Daniel Braun}
\affiliation{Laboratoire de Physique Th\'eorique, IRSAMC, UMR5152 du CNRS,
  Universit\'e 
  Toulouse Paul Sabatier, 31062 Toulouse Cedex 4, FRANCE} 
\date{\today}
\begin{abstract}
Two non--interacting quantum systems which couple
to a common environment with many degrees of freedom initially in thermal
equilibrium can become entangled
due to the indirect 
interaction mediated through this heat bath. I examine here the dynamics of
reservoir induced entanglement
for a heat bath consisting of a thermal electro--magnetic radiation field,
such as black body radiation or the cosmic microwave background, and show
how the effect can be understood as result of an effective induced
interaction.  
\end{abstract}
\maketitle

\section{Introduction}
About two decades of research in quantum
information have led to the picture of quantum entanglement as
a precious resource that plays a key role in processing information more
securely and more efficiently than classically possible
\cite{Bennett00}. Entanglement is necessary for quantum teleportation
\cite{Bennett93} and the exponential acceleration of quantum algorithms
\cite{Jozsa02}, and it allows for secure quantum key
distribution \cite{Ekert91}. 
Recent experimental demonstrations of quantum 
teleportation
\cite{Bouwmeester97,Nielsen98,Ursin04,Zhao04,Barrett04,Riebe04} and
small scale quantum computation \cite{Vandersypen01} confirm this picture.
Creating and manipulating
entanglement in a controlled way remains a challenge, as
environmentally induced decoherence tends to rapidly destroy
entanglement. On the other hand, it has been shown that the coupling of
two quantum systems to 
a common heat bath can also {\em create} substantial (mixed state) 
entanglement. This effect was originally
demonstrated in the framework of an exactly solvable model \cite{Braun02}, and
has 
been confirmed by perturbative calculations in the Markovian regime
\cite{Benatti03,Jacobczyk02}. Cirone et al.~have unveiled a connection to the
Casimir-Polder interaction \cite{Cirone04}, and Oh and Kim have shown by a
renormalization group analysis how
the effect can be understood through 
an effective induced interaction between the quantum systems {\em via} the
common heat bath \cite{Oh04}.

Heat baths are ubiquitous, and one might therefore wonder, if
``reservoir induced entanglement'' (RIE) \cite{Chen05}, i.e.~creation of
entanglement through coupling to a common heatbath is   
common place and over what distances and with what time dependence it might
arise. For example, all quantum systems containing 
charged particles couple to the ambient thermal electro--magnetic radiation,
i.e.~the black body radiation (BBR) in a laboratory or cryostat at the
corresponding temperatures, and even in free space there is still the
cosmic micro--wave background (CMB), a 
basically perfect black body radiation at an absolute 
temperature  
$T=2.728\pm 0.004$ K  that fills the entire known
universe \cite{Bennett03}.   Do these heat baths induce
entanglement between remote quantum systems? 

In this paper I show that BBR effectively constitutes two different heat
baths which couple differently to the couple Alice--Bob, and whose effects
largely cancel when it comes to entanglement creation.  Entanglement is
therefore created only very slowly, far behind the light--cone, and the
entanglement 
created oscillates as a function 
of time. The 
entanglement can be close to perfect, but the first maximum
entanglement arrives only after a time $t_1$ which scales like the 3rd power
of the distance $R$ between the two quantum systems,
\begin{equation} \label{tmax}
t_1\simeq\frac{\pi}{2\alpha_0}\frac{R^2}{d^2}\frac{R}{c_0}
\end{equation}
where $\alpha_0\simeq 1/137$ and $c_0$ are the  fine structure constant
and the speed of light in vacuum, 
respectively, and $d$ denotes the dipole moment of the quantum system
divided by the electron charge. The slow creation of entanglement 
limits the 
distance over which it can be created before competing decoherence processes
set in. 

I will elucidate the role of the high-frequency, far off-resonant modes
of the heat bath, and discuss the temperature dependence of the phenomena.

\section{Theoretical Framework and Physical Systems}
Let us consider two identical quantum systems A and B which couple to the
thermal electromagnetic radiation field in open space, as exemplified by the
CMB, 
but which do not interact directly. We assume that A and B can be
approximated as two state systems with states $|0\rangle$ and 
$|1\rangle$. 
It turns out that the standard quantum optics approach of rotating wave
approximation and Markovian Master equations based on Fermi Golden Rule
rates (i.e. second order perturbation 
theory in 
the atom--field coupling constants) is not fully adequate for describing
RIE. Firstly, the explicit dependence of the time scale on which RIE is
produced on the cut--off frequency of the heat bath found in \cite{Braun02}
hints to the 
importance of non--resonant modes, which invalidates the rotating wave
approximation. One might argue that in \cite{Braun02} non--resonant modes
came into play because degenerate energy levels were considered, but even
for finite level spacing $\Delta$ the high--frequency modes should be
relevant for 
times $t\ll\Delta$. Note that the rotating wave approximation was also avoided
in \cite{Reznik03,Oh04}.  

Secondly, there is evidence that for the specific heat bath and quantum
systems to be discussed here RIE is an effect that arises only at fourth
order in the coupling constants \cite{Braun05}. Both issues combined
call for a fourth order 
calculation without rotating wave approximation, which makes the theory very
heavy. On the other hand, the problem can be solved
exactly and with less effort in the case of degenerate energy levels
\cite{Braun01,Braun02}.  This approach will be followed here.
 
From an experimental point of view, two--state systems with exactly
degenerate energy levels are hard to find. However, it should be kept in
mind that a finite level spacing just introduces an upper limit to
the time for which the present theoretical analysis is applicable: for any
experiment terminated within a time $t\ll 1/\Delta$ the system
hamiltonian of the two--state systems A and B can be neglected
\cite{Braun01}. Whether or not entanglement can still be produced beyond
this time is an interesting experimental question. 

Double
quantum dots (DQDs) seem to be a promising candidate, and the following
analysis will be geared specifically towards these systems.
Recently, the coherent  
manipulation of two states $|0\rangle$ and $|1\rangle$ located in the two 
wells of a single such device, as well as state preparation and state
measurement were demonstrated; coherence times of the order 1-10ns
 were achieved \cite{Hayashi03,Petta04}. 
The energy barrier between the two wells as well as the energies of
$|0\rangle$ and $|1\rangle$ in a DQD can be tuned with
the help of gate voltages and the barrier can be made 
very high after the preparation of a superposition, such that
$|0\rangle$ and $|1\rangle$ become to good approximation
degenerate eigenstates of the DQD hamiltonians. The question of whether or not
the two state approximation still holds when we have to consider
high--frequency modes might be addressed eventually experimentally by
trying to shield each DQD with its own superconducting cavity. Modes with
frequency higher than the superconducting gap will be absorbed and one might
thus envisage to control the cut--off frequency of the heat bath. 
Otherwise a cut--off of the order $\hbar\omega\simeq 1eV$ arises
naturally due to the band--gap of the 
semi--conductor material in which the DQDs are embedded; e.m.~waves with
higher frequencies get absorbed in the semi--conductor. 
A disadvantage of the  DQDs are the competing intrinsic decoherence
mechanisms such as phonon scattering
\cite{Vorojtsov04} and fluctuating electric fields \cite{Itakura03} other
than those from the BBR, which will be neglected in the following analysis.

As for the heat bath, I will consider specifically BBR with periodic
boundary conditions. One might wonder if the mode structure (and thus
the boundary conditions) make a difference, as each mode couples to two
spatially separated quantum systems. For the geometrical situation
considered below, and within 
dipole approximation of the coupling, it turns out that a box--shaped cavity
with 
perfectly conducting walls leads, in the limit of infinite volume and fixed
distance $R$, to the same interaction 
hamiltonian, and one can thus read in the following CMB with periodic
boundary conditions or BBR in a
box--shaped cavity interchangeably. 

\section{Double Quantum Dots interacting with BBR}
Let us arrange the two DQDs such that the axes of the dots are aligned with
the vector joining them, designated as $z$ axis in the following.  
The position operators of an electron in dot 1 and 2 have the matrix
elements $\langle 
0|z_{1,2}|0\rangle=-d/2=-\langle 
1|z_{1,2}|1\rangle$. All other matrix elements of $z_{1,2}$ vanish to
very good approximation due to the 
exponentially small  overlap of the states $|0\rangle$ and
$|1\rangle$, and so do the matrix elements of the other electron
coordinates $x_{1,2}$  and $y_{1,2}$ for an assumed even parity of the
ground state wave functions.
 
For describing the BBR I use a
slightly unconventional representation of the electro--magnetic field, which
is very useful for the exact treatment for vanishing level
spacing (or in general, if one does not use the rotating wave
approximation). In fact, the  
BBR can be considered as two
independent heat baths (see appendix \ref{appI}), one containing cos--waves,
the other 
sin--waves, with an electric field operator
\begin{eqnarray}
\bE(\br,t)&=&\sum_{\bk,\alpha}'\sqrt{\frac{2}{\epsilon_0V}}
\omega_{k}\beps_{\bk \alpha}
\left(
Q_{\bk \alpha 1}\cos\bk \br+
Q_{\bk \alpha 2}\sin\bk \br\right)\,,\label{efield}
\end{eqnarray}
and canonical coordinate operators $Q_{\bk\alpha\nu}$ of the 
harmonic oscillators corresponding to the different
electro--magnetic field modes.
The $\bk$ are quantized wave vectors ($k_i=2\pi n_i/L$,
$n_i\in{\cal Z}$ for $i=x,y,z$, for quantization in a volume $L^3$ with
periodic boundary 
conditions), $\alpha=1,2$ counts
polarizations, $\nu$ distinguishes the cos--waves ($\nu=1$) from the
sin--waves ($\nu=2$), and the $'$  
on the sum means summation 
restricted to  $k_x>0$. The
free field hamiltonian reads  
\begin{eqnarray}
H_{\rm
  bath}&=&\frac{1}{2}\sum_{\bk}'\sum_{\alpha=1,2}\sum_{\nu=1,2}\left(P_{\bk \alpha\nu}^2+\omega_{k}^2Q_{\bk \alpha\nu}^2\right)\,,\label{Hbath}
\end{eqnarray}
with $\omega_k=c_0|\bk|\equiv c_0k$.

Placing the two DQDs
at positions ${\bf R}/2$ and $-{\bf R}/2$ (${\bf R}=(0,0,R)$) we obtain in
dipole approximation the coupling hamiltonian 
\begin{eqnarray}
H_{\rm
  int}&=&\sum_{\bk}'\sum_{\alpha=1,2}\left(\left(\sigma_{z1}+\sigma_{z2}\right)g_{\bk
    \alpha 1}Q_{\bk
    \alpha 1}+\left(\sigma_{z1}-\sigma_{z2}\right)g_{\bk
    \alpha 2}Q_{\bk
    \alpha 2}\right)\,,\label{Hint}
\end{eqnarray}
written in terms of Pauli matrices $\sigma_{z1}$, $\sigma_{z2}$ in the
basis $(|0\rangle,|1\rangle)$ of dots 1 and 2.  
The coupling coefficients $g_{\bk \alpha\nu}$ are given by
\begin{eqnarray}
g_{\bk\alpha
1}&=&\frac{ed}{2}\sqrt{\frac{2}{V\epsilon_0}}\omega_k\beps_{\bk\alpha}\cdot\left(
\begin{array}{c}
0\\0\\1
\end{array}
\right)\cos\left(\frac{\bk \bR}{2}\right)\,,\label{gka}
\end{eqnarray}
and the same equation holds for $g_{\bk\alpha 2}$ up to the change
$\cos\to\sin$. 

In the regime of energy degenerate states $|0\rangle$ and $|1\rangle$
discussed above, the time
evolution of each dot due to its own system hamiltonian can be
neglected, and the total hamiltonian is thus simply $H=H_{\rm
  int}+H_{\rm bath}$\cite{Braun01,Braun02}. The interaction
hamiltonian represents a 
generalization of the situation considered in \cite{Braun01}, in the
sense that each mode of the BBR couples through a different ``system
coupling agent'' $S_{\bk\alpha\nu}=g_{\bk
  \alpha\nu}(\sigma_{z1}-(-1)^\nu\sigma_{z2})$. Note, however,
that all coupling agents commute with each other. This allows to
generalize the time evolution derived in \cite{Braun01,Braun02} of the
reduced density matrix 
describing the two DQDs alone to 
\begin{eqnarray}
\langle
s|\rho(t,t_0)|s'\rangle&=&\exp\Big\{-\sum_{\bk,\alpha,\nu}'\Big[
\left(
\lambda_{s,\bk\alpha\nu}-\lambda_{s',\bk\alpha\nu}
\right)^2f(k)\\\nonumber
&&-\ri\left(
\lambda_{s,\bk\alpha\nu}^2-\lambda_{s',\bk\alpha\nu}^2
\right)
\varphi(k) 
\Big]
\Big\}\langle
s|\rho(0,t_0)|s'\rangle,
\end{eqnarray}
where $|s\rangle$ is one of the four states $|00\rangle$,
$|01\rangle$, $|10\rangle$, or $|11\rangle$, and
$\lambda_{s,\bk\alpha\nu}$ are the corresponding eigenvalues of
$S_{\bk\alpha\nu}$. It was assumed that the bath is initially in thermal
equilibrium, and that the total initial density matrix factorizes into a
system part and a bath part. The dependence on the time of
travel of a light signal between the two 
DQDs, $t_0=R/c_0$ with $R=|\bR|$, will appear below.

For the arrangement of the DQDs
described above, only one polarization direction ($\alpha=1$) contributes,
with $\beps_{\bk 1}=\beps_\theta$ in polar coordinates for $\bk$. With the
abbreviation $G_{k}=-\frac{ed}{2}\sqrt{\frac{2}{V\epsilon_0}}\omega_k
\sin\theta$ the relevant eigenvalues are $\lambda_{00,\bk 11}=2\cos(k R
\cos\theta/2)G_k=-\lambda_{11,\bk 11}$ and $\lambda_{01,\bk 12}=2\sin(k R
\cos\theta/2)G_k=-\lambda_{10,\bk 12}$. All other eigenvalues vanish.  The
functions $f(k)$ and 
$\varphi(k)$ depend on $k=|{\bf k}|$ as ($\beta=1/(\kb T)$)
\begin{eqnarray}
f(k)&=&\coth\left(\frac{\beta\hbar\omega_{k}}{2}\right)\frac{1-\cos\omega_{k}t}{2\hbar\omega_{k}^3}\mbox{, }
\varphi(k)=\frac{1}{2\hbar\omega_{k}^2}\left(t-\frac{\sin\omega_{k}t}{\omega_{k}}\right)\,.
\end{eqnarray}
Transforming the sums over modes ${\bk}$ into integrals for the limit
of large $L$, we find 
\begin{eqnarray}
\langle s_1|\rho(t,t_0)|s_2\rangle&=&\exp\Bigg(
-A\Big(f_1(t,t_0) C_{s_1,s_2}+f_2(t,t_0) S_{s_1,s_2}-\ri\big(   \varphi_1(t,t_0) \tilde{C}_{s_1,s_2}+\varphi_2(t,t_0) \tilde{S}_{s_1,s_2}      \big)
\Big)
\Bigg)\langle
s_1|\rho(0,t_0)|s_2\rangle\,,\label{rhofinal}
\end{eqnarray}
with 
\begin{eqnarray}
S=\left(
\begin{array}{cccc}
0&1&1&0\\
1&0&4&1\\
1&4&0&1\\
0&1&1&0
\end{array}
\right)\mbox{, }
C=\left(
\begin{array}{cccc}
0&1&1&4\\
1&0&0&1\\
1&0&0&1\\
4&1&1&0
\end{array}
\right)\mbox{, }
\tilde{S}=\left(
\begin{array}{cccc}
0&-1&-1&0\\
1&0&0&1\\
1&0&0&1\\
0&-1&-1&0
\end{array}
\right)\mbox{, }\tilde{C}=-\tilde{S}\,,
\end{eqnarray}
in the basis $|00\rangle$,
$|01\rangle$, $|10\rangle$, $|11\rangle$, and $A=\alpha_0 \frac{d^2}{\pi c_0^2\tau^2}$, where
$\alpha_0=e^2/4\pi\epsilon_0\hbar c_0\simeq
1/137$ is the fine--structure constant.  In the following, both $t$ and
$t_0$ will 
be expressed in units of the thermal time $\tau=\beta\hbar$ for all
finite temperatures.
The functions $f_{1,2}(t,t_0)$ and $\varphi_{1,2}(t,t_0)$ are then given by
\begin{eqnarray}
f_\nu(t,t_0)&=&\int_{0}^{y_{max}}dy\,y\coth(y/2)\left(1-\cos\left(yt\right)\right)\left(\frac{1}{3}+(-1)^\nu\left(\frac{\cos(yt_0)}{(yt_0)^2}-\frac{\sin(yt_0)}{(yt_0)^3}\right)\right)\label{fs}\\
\varphi_\nu(t,t_0)&=&\int_{0}^{y_{max}}dy\,y\left(yt-\sin\left(yt\right)\right)\left(\frac{1}{3}+(-1)^\nu\left(\frac{\cos(yt_0)}{(yt_0)^2}-\frac{\sin(yt_0)}{(yt_0)^3}\right)\right)\label{phis}\,,
\end{eqnarray}
The integrals extend in principle from zero to infinity, but a UV  cut--off
$y_{max}=\omega_{max}\tau$ is needed to regularize them.

Due to $\tilde{C}=-\tilde{S}$, the final density matrix depends only
on the phase difference $\varphi_-(t)=\varphi_1(t)-\varphi_2(t)$, which can
be written in closed form as
\begin{eqnarray}
\varphi_-(t,t_0)&=&\frac{t}{t_0^3}\left(-2\sin\left(y_{max}t_0\right)+\rSi\left(y_{max}(t-t_0)\right)+2{\rSi}\left(y_{max}t_0\right)-\rSi\left(y_{max}(t+t_0)\right)\right)\nonumber\\
&&+\frac{2}{y_{max}t_0^3}\sin\left(y_{max}t \right)\sin\left(y_{max}t_0 \right)\,.\label{phim}
\end{eqnarray}
Note that $\varphi_-(t,t_0)$ remains finite for $y_{max}\to\infty$. For
large $t$ and $t_0$, $\varphi_-(t,t_0)$ increases proportional to $t$ and
decays as $1/t_0^3$. The functions $f_\nu(t,t_0)$ on the other hand scale
like $y_{max}^2$ with the cut--off, which gives a physical significance
to the high--frequency modes. 
The influence of the form of the cut--off function will be
examined 
below. For the moment we assume the simplest form, a sharp cut--off at
$\omega=\omega_{max}$.   

\section{Reservoir Induced Entanglement}
\begin{figure}
\includegraphics{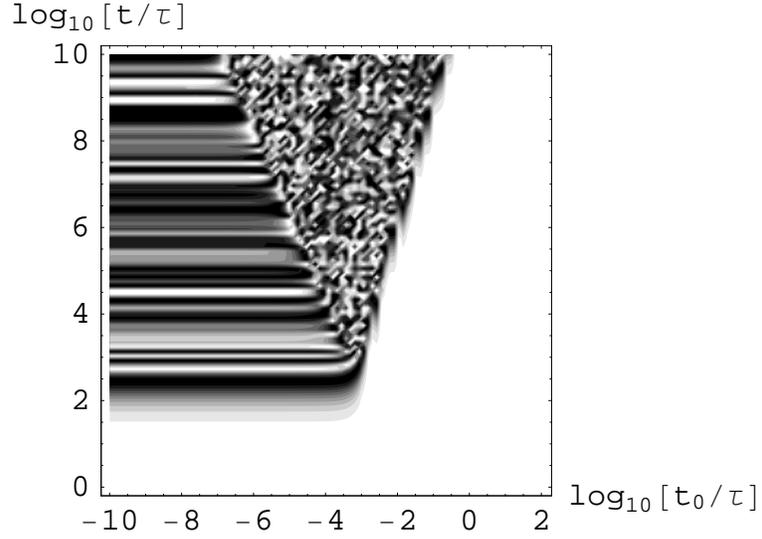}
\caption{Entanglement of formation  $E$ for two initially
not entangled DQDs with $d=10$nm coupled to the
CMB at $T=2.73$K as a function of $\log_{10}(t_0/\tau)$ and
$\log_{10}(t/\tau)$, $\tau=\beta\hbar$. Black means
perfect entanglement, 
$E=1$, white no 
entanglement $E=0$. Entanglement is created only for $t/\tau\gtrsim
10^{12}(t_0/\tau)^3$.}\label{fig1}      
\end{figure}

\begin{figure}
\includegraphics{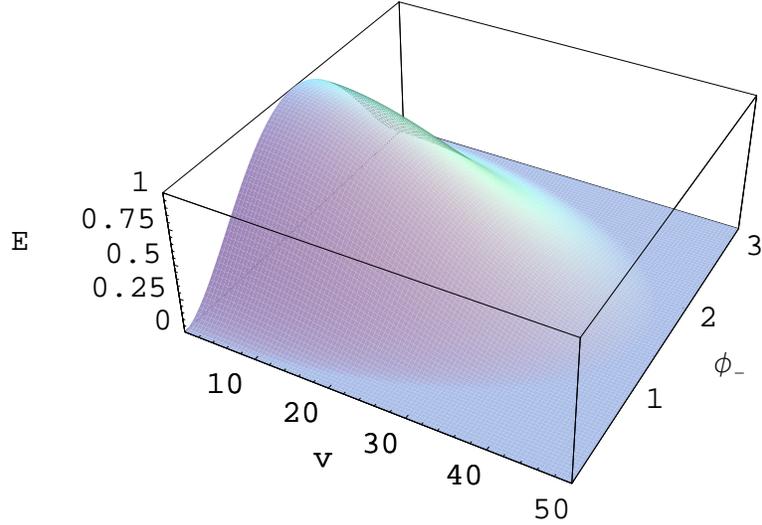} 
\caption{Entanglement of formation for $t\gg t_0 \gg \tau$ as function of
  $v=\omega_{max}d/c_0$ and the phase $\varphi_-$. The latter increases
  linearly with time $t$ leading to entanglement oscillations, and decays as
  $t_0^{-3}=(R/c_0)^{-3}$, which makes the entanglement creation very slow
  for large distances $R$.}\label{fig2}    
\end{figure}

I have evaluated the entanglement of formation $E(\rho(t,t_0))$ for an
initial state in the form of a pure product state,
$(|0\rangle+|1\rangle)\otimes (|0\rangle+|1\rangle)/2$,
i.e.~$E(\rho(t,t_0))=0$ using 
Wootter's formula \cite{Wootters98} and by
numerically integrating the remaining expressions for
$f_{1,2}(t,t_0)$. Alternatively, one may approximate $\coth\simeq 1$ for
$y_{max}\gg 1$, which allows for an analytical solution also for $f_\nu$. The
result for $T=2.73$K and $d=10$nm is shown in Fig.\ref{fig1}. Clearly, there
is no 
entanglement for space--like separated points, $t<t_0$. However, almost
perfect mixed state entanglement 
arises for $t>t_0$, starting at around $t\simeq 100\tau$. The entanglement
oscillates rapidly as function of $t$, and vanishes for
$t/\tau< c (t_0/\tau)^3$, where $c$ is a constant of order $10^{12}$ in
agreement with eq.(\ref{tmax}). 
This behavior can be understood by considering the long time limit
$t\gg t_0\gg\tau$, which leads to 
$f_{1,2}\simeq y_{max}^2/6$ for
$y_{max}\gg 1$. In this case the final density matrix depends only on two
variables, 
$v=d\omega_{max}/c_0$ and $\phi_-=A\varphi_-=\phi_-(t,t_0)$, with 
\begin{eqnarray}
\langle
s|\rho(v,\phi_-)|s'\rangle&\simeq&\exp\left(-\frac{\alpha_0}{12\pi}v^2(S_{s,s'}+C_{s,s'})+\ri\phi_-\tilde{C}_{s,s'}\right)\langle
s|\rho(v,\phi_-(0,t_0))|s'\rangle\,.
\end{eqnarray}
Fig.\ref{fig2} shows the corresponding entanglement. The function is
$\pi$--periodic in $\phi_-$ with a first maximum at $\phi_-=\pi/2$, which
leads to eq.(\ref{tmax}) for the first maximum entanglement. The scaling of
$t_{1}$ with $R^3$ makes the entanglement production extremely slow:
during the 
lifetime of the universe, it would have reached a distance of only about
8.4km for $d=1\mu$m. For the same dipole moment, a
distance of about 52$\mu$m should be reached during a coherence
time of the order of 100ns. The maximum amount of entanglement 
 as well as the time to first finite entanglement creation is controlled by
$v$, which has to be smaller than about 50. One should therefore try to have
 a large $d$ for a large maximum distance and a small $\omega_{max}$  to get
 large maximum entanglement, whereas the temperature becomes irrelevant in
this regime ($\hbar\omega_{max}\gg \kb T$). Note that $\tau$ cancels in
the expression 
for $\phi_-$ for large $t,t_0$, such that also $t_1$ becomes 
independent of temperature. 

\noindent

\section{The effective interaction}
The physical origin of the entanglement generation can be understood as
arising from an effective interaction $H_{\rm eff}$ mediated through the
coupling to the common bath \cite{Oh04}. This idea can be made quantitative
by observing 
that a Hamiltonian of the type
\begin{equation} \label{Hk}
H=\sum_k\hbar\omega_ka_k^\dagger a_k+\hbar\sum_kg_k(S_ka_k+
S_k^\dagger a_k^\dagger)  
\end{equation}
where $S_k$ is an arbitrary operator acting on the Hilbert space of the
system (i.e. here the qubits of Alice and Bob) leads to the time evolution
operator \cite{Mahan90} 
\begin{equation} \label{UMa}
\re^{-\ri/\hbar H
  t}=D(-\sum_k\frac{g_k}{\omega_k}S_k^\dagger)\re^{-\ri/\hbar (H_{\rm
  bath}+H_{\rm eff})t}D(\sum_k\frac{g_k}{\omega_k}S_k^\dagger)\,,
\end{equation}
with the shift operator $D(\sum_k\frac{g_k}{\omega_k}S_k^\dagger)=\exp\left(
\sum_k(g_k/\omega_k)(S_K^\dagger a_k^\dagger-S_ka_k)\right)$, bath hamiltonian $H_{\rm
  bath}=\sum_k\hbar\omega_ka_k^\dagger a_k$, and the effective interaction 
\begin{equation} \label{Heff}
H_{\rm eff}=-\sum_k\frac{g_k^2}{\omega_k}S_kS_k^\dagger\,.
\end{equation}
It is very instructive to calculate this interaction explicitly for the
example at hand. To that end we revert momentarily to the standard
expression of the electric field, eq.(\ref{E}), and allow for arbitrary
orientation of the DQDs. If we denote the orientations by unit vectors
$\hat{u}_i$ and observe that the index $k$ in eq.(\ref{Hk}) stands for wave
vector and polarization
$(\bk,\alpha)$, we have
$S_{\bk\alpha}=\sum_{i=1,2}g_{\bk\alpha}^{(i)}\sigma_{zi}\re^{\ri\bk\br_i}$
with coupling constants  
\begin{equation} \label{gk}
g_{\bk\alpha}^{(i)}=\bd_i\cdot\epsilon_{\bk\alpha}\sqrt{\frac{c_0|\bk|}{2\epsilon_0\hbar
V}}\,, 
\end{equation}
with $\bd_i=(ed/2)\hat{u}_i$. The evaluation of the sum
over all modes in eq.(\ref{Heff}) leads, in the limit of continuous $\bk$
and infinite cut--off frequency, to 
\begin{equation} \label{Heff1}
H_{\rm
  eff}=\frac{\bd_1\cdot\bd_2-3(\bd_1\cdot\hat{r})(\bd_2\cdot\hat{r})}{4\pi\epsilon_0r^3}\sigma_{z1}\sigma_{z2}\,,
\end{equation}
i.e. a dipole--dipole interaction. This explains the $1/r^3$
dependence of the phase $\varphi_-(t,t_0)$ and its proportionality to $t$
for large times. Of course, the electric field can not mediate an
instantaneous interaction, as eq.(\ref{Heff1}) might suggest. The full time
dependence must take into account also the free bath hamiltonian, and
$H_{\rm bath}+H_{\rm eff}$ together lead indeed to the correct retardation
behavior. This is seen from eq.(\ref{phim}) when we take the limit
$\omega_{\max}\to\infty$. The first oscillating term $\sin(y_{max}t_0)$
arises from the sharp cut-off and will average to zero for a smoother
cut-off. The last term vanishes for $y_{max}\to\infty$, and the remaining
sin-integral functions conspire to a Heaviside  theta--function on the
light--cone, $\pi\theta(t/t_0-1)$, with the consequence that no
entanglement can be created faster than the speed of light with this effect. 
The matrix $\tilde{C}$ in eq.(\ref{rhofinal}) is seen to arise from the
difference of eigenvalues of $\sigma_{z1}\sigma_{z2}$ defined by the indices
of the density matrix. 

One might be tempted to think at this point that the entanglement generation
is trivial in the sense that the hamiltonian for the two DQDs
and the heat bath just amounts to a fancy reformulation of the dipole--dipole
interaction between the DQDs initially assumed non--interacting. This is,
however, not the case: First of all, as can be seen from eq.(\ref{UMa}) the
effective interaction is not the only term that determines the dynamics of
the entanglement generation. Rather it is supplemented by two shifts in the
harmonic oscillators, which depend on the state of the two DQDs. The reduced
overlaps of the harmonic oscillators are responsible for the decoherence
quantified by the functions $f_1(t,t_0)$ and $f_2(t,t_0)$. Thus, there needs
to be a balance between the effective interaction induced and the
decoherence due to state dependent coupling to the heat bath, and this
balance is made quantitative by the functions $f_1(t,t_0)$, $f_2(t,t_0)$,
and $\varphi_-(t,t_0)$. The effect of decoherence (and possibly the
retardation) would be overlooked, if one
started directly with a hamiltonian containing the dipole--dipole
interaction. 
 
Secondly, the heat bath is initially in thermal equilibrium, and the thermal
noise reduces the amount of entanglement additionally compared to the $T=0$
case. So the nature of the environment as a heat bath is important, and not
just the fact that it induces a well--known interaction. 

\begin{figure}
\includegraphics{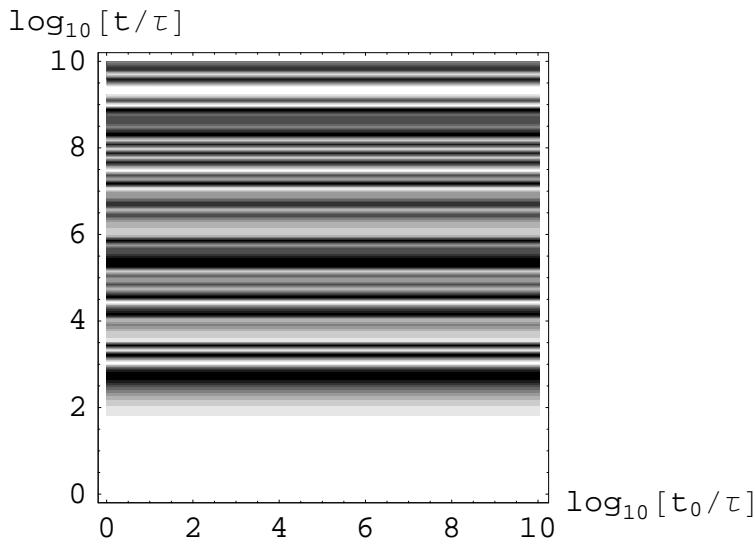} 
\caption{Entanglement of formation  $E$ for a hypothetical 
coupling to the cos--waves only. Entanglement production is basically
independent of $R$ in the interval shown, and
thus possible  quasi-instantaneously ($t<t_0$). Same parameters and grey--scale
code as in 
Fig.\ref{fig1}}\label{figsym}     
\end{figure}

One might wonder whether one can speed up the entanglement generation by
reservoir engineering. One obvious attempt would be to 
selectively supress the coupling to one of the two heat baths, say the
sin--waves. In fact, the calculation shows
that for a complete supression of the coupling to the sin--waves the
entanglement  
becomes basically independent of $R$ and entanglement would be created
quasi--instantaneously, Fig.\ref{figsym}. The
difference arises from the fact that 
$\varphi_-=\varphi_1-\varphi_2$, a 
phase difference accumulated from the couplings to the $\cos$-- and $\sin$--
waves, is replaced by a single phase from the $\cos$--waves alone. The
latter contains a large distance independent term which normally cancels a
corresponding term from the $\sin$--waves. Without the $\sin$--waves
($\varphi_2=0$) this term
leads to a very rapid, distance independent growth of $\varphi_-$ and
therefore to the quasi instantaneous entanglement creation. 
We conclude that the two independent
heat baths counteract each other
when it comes to entanglement creation, and this leads to much slower
entanglement creation based on the small remaining, distance dependent
phase accumulation. 

\begin{figure}
\includegraphics{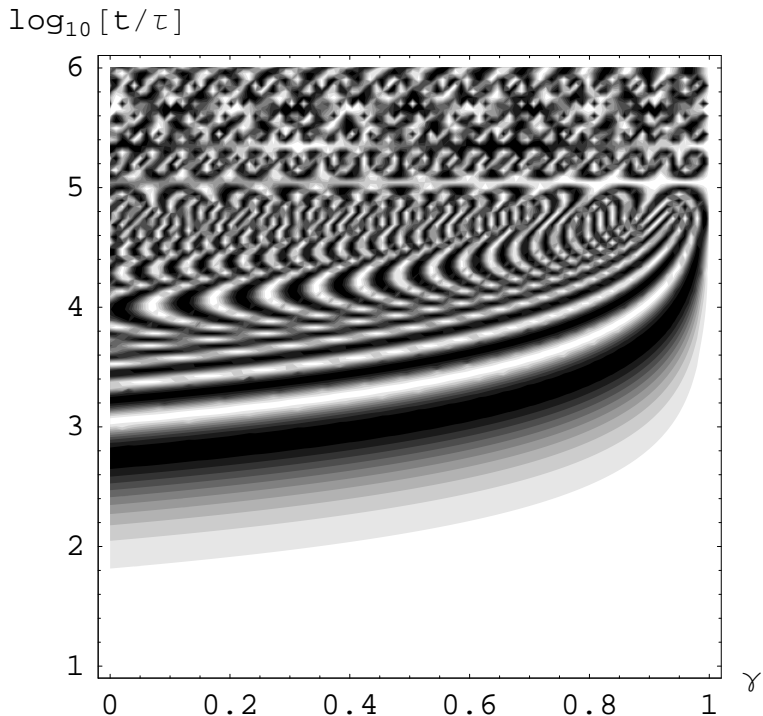} 
\caption{Entanglement of formation for $t_0 =10^6\tau$ as function of
  $\gamma$ and $\log_{10}(t/\tau)$. The parameter $\gamma$ measures the
coupling strength to the sin--waves (0 no coupling, 1 full coupling). Already
values of $\gamma$ only slightly smaller than 1 would lead to
  entanglement creation before the light cone 
($t<t_0$). Same parameters and grey--scale code as in
  Fig.\ref{fig1}.}\label{fig3} 
\end{figure}

While a comlete suppression of the sin--waves (or the coupling to them)
might be illusive, even a very 
small suppression (as one might imagine in a cavity by using a thin,
uncharged wire) would 
lead quickly to entanglement
creation faster than the speed of light. Fig.\ref{fig3} shows the
entanglement for $t_0=10^6\tau$ as 
function of $\gamma$ and $\log_{10}(t/\tau)$, where $0\le\gamma\le 1$ measures
the relative coupling strength to the sin--waves ($f_2(t,t_0)\to\gamma
f_2(t,t_0)$ and $\varphi_2(t,t_0)\to\gamma \varphi_2(t,t_0)$). The smaller
$\gamma$, the faster the entanglement arises, but up to $\gamma$ very close
to unity, almost perfect entanglement is created for $t<t_0$. 
\noindent

However, causality leads to strong restrictions of what should be possible
in this respect: also in a classical field
theory the retardation of the dipole--dipole interaction would be modified,
and a electromagnetic interaction spreading faster than the speed of light
is certainly not possible, regardless of whatever arrangement of conductors
one might come up with in order to supress certain modes. 

In \cite{Reznik03,Reznik05} it was proposed that two atoms can get entangled by
coupling them for a time $t<t_0$ to the vacuum of a massless scalar
relativistic field (thus explicitly avoiding the effects of any effective
induced interaction), and an experiment was proposed in  \cite{Retzker04}  to
demonstrate the corresponding effect in ion-traps, namely the entangling of
the internal degrees of freedom of two ions in an ion-string faster than the
speed of sound in the ion-string. The vacuum case can be retrieved from the
calculation presented here by setting $T=0$, and replacing the $\coth$  
function in
eq.(\ref{fs}) by unity. Since $\tau$ diverges then, it
is more reasonable to use directly $t_0$ as time scale. This is achieved
formally by substituting $y=z\tau/t_0$ in eqs.(\ref{fs},\ref{phis})
whereupon the functions depend only on $t/t_0$ up to an additional prefactor
$\tau^2/t_0^2$. The latter combines with the prefactor $A$ and reads then
$\alpha_0 d^2/(\pi c_0^2t_0^2)$. The cut-off $y_{max}$ is replaced by
$z_{max}=\omega_{max}t_0$. However, nothing changes about the fact that the
  function $\varphi_-$ is proportional to a Heaviside-theta function centered
  on the light-cone. Thus, no phase accumulation is possible for $t<t_0$,
  unless one uses a finite cut-off. In this case the Heaviside theta
  function gets ``softened'', which is physically plausible as now the
  heat-bath 
  does not contain sufficiently small wave-lengths anymore to precisely locate
  Bob's DQD. But even if in such a case one was to operate slightly before the
  light-cone, 
  $t\lesssim t_0$, one would still have to overcome the smallness of the
  prefactor $\alpha_0 d^2/(\pi c_0^2t_0^2)$. This seems to be excluded for
  atoms, as the distance between Alice and Bob would have to become smaller
  than the length attached to the atomic dipole moment $d$, in which case
  the whole dipole approximation breaks down. In DQDs one
  might hope to create states $|0\rangle$ and $|1\rangle$ with large amounts
  of charge, thus increasing artificially the dipole moments, but even for
  ten excess electrons in one well  
  the DQDs would have to be so close that the dipole approximation becomes
  doubtful. Thus, within the above theoretical framework of fixed 
  DQDs with two degenerate energy levels dipole--coupled to BBR, there seems to
  be not much room 
  for substantially 
  entangling the atoms faster than the speed of light by coupling them
  for a time $t<t_0$ to the 
  BBR, even at $T=0$. 

\section{Cut--off Dependence} 
As the function $\varphi_-$ remains finite for
$\omega_{max}\to\infty$, a  different cut--off
function has little influence on $\varphi_-$ (besides the eventual removal
of the oscillating terms mentioned earlier) and therefore on the
speed of entanglement generation. However, since $f_{1,2}$ diverge for
$y_{max}\to\infty$, the cut--off function can change the maximum
amount of entanglement. Let us suppose that the cut--off function to
be used in eq.(\ref{fs}) is equal to unity for $y<y_{max}$ and equal to a
function $C(y)$ for $y\ge y_{max}$. For $y_{max}\gg 1$, the $\coth(y/2)$ can
be replaced by unity without impacting the following scaling
arguments. Also, the strongest diverging $t_0$--dependent part is given by 
\begin{equation} \label{t0dep}
\int_{y_{max}}^\infty dy\,C(y) \frac{\cos(yt_0)}{yt_0^2}\,,
\end{equation}
which even for $C(y)=1$ remains finite.
The only remaining question is then, if a different cut--off function
changes significantly the $t$--dependent behavior of $f_{1,2}$, and in
particular, if for $t\to\infty$ $f_{1,2}$ always remain finite. To answer
this question observe that for $0\le C(y)\le 1$,
\begin{equation} \label{absch}
0\le \frac{1}{3}\int_{y_{max}}^\infty dy\,C(y) y (1-\cos(yt))\le
\frac{2}{3}\int_{y_{max}}^\infty dy\,C(y) y \,. 
\end{equation}
Thus, if the cut--off function decays fast enough to make the integral
finite, it will be finite for all times. More specifically, if $C(y)\propto
1/y^p$, $p$ needs to be larger than two, and the upper bound in
(\ref{absch}) is then of the order $~y_{max}^{2-p}$, which is completely
negligible compared to the dominating $y_{max}^2$ behavior from
$y<y_{max}$. Therefore, the results obtained, in particular the entanglement
for large $t$, are robust against a change of the cut--off function.


\section{Conclusions} \label{secConcl}
The CMB or the black body radiation (BBR) in a box--shaped cavity can be
regarded 
as two independent heat baths, which couple differently to two spatially
separated DQDs. Two 
spatially separated, non--interacting quantum systems can get entangled by
interacting with the CMB or the same BBR. 
However, the 
effects of both heat baths cancel to large extent. As a result, entanglement
is created only very slowly,
with a first maximum entanglement arriving at a time $t_1$ that scales  like
the 3rd power of the distance $R$ between
the quantum systems. The effect can be understood as originating from an
effectively induced dipole-dipole interaction, and consequently is retarded
by the propagation of a light signal. 

{\em Acknowledgment:} I thank Jaewan Kim and Sangchul Oh for useful
  discussions, and Benni Reznik for useful correspondance.

\section{Appendix: The Electro--magnetic Field} \label{appI}
I derive here the expression for the electric field operator,
eq.(\ref{efield}). Starting point is the standard representation of the
quantum operator of the electric field for
periodic boundary conditions \cite{Scully97},
\begin{equation} \label{E}
\bE(\br,t)=\sum_{\bk,\alpha}\beps_{\bk\alpha}{\cal E}_k\left(a_{\bk\alpha}\re^{-\ri
(\omega_kt-\bk\br)}+a^\dagger_{\bk\alpha}\re^{\ri
(\omega_kt-\bk\br)}\right)\,,
\end{equation}
with ${\cal E}_k=\sqrt{\hbar\omega_k/(2\epsilon_0V)}$, unit polarization
vectors 
$\beps_{\bk\alpha}$ ($\alpha=1,2$), and the creation and annihilation
operators $a^\dagger_{\bk\alpha}$ and $a_{\bk\alpha}$, respectively. We
express these operators in terms of canonical coordinate and momentum
operators $q_{\bk\alpha}$ and $p_{\bk\alpha}$, respectively, to obtain
\begin{equation} \label{E2}
\bE(\br,t)=\sum_{\bk,\alpha}\beps_{\bk\alpha}\frac{1}{\sqrt{\epsilon_0V}}\left(\omega_{k}q_{\bk\alpha}\cos(\bk\br-\omega_k
t)-p_{\bk\alpha}\sin(\bk\br-\omega_k t)
\right)\,.
\end{equation}
We then
split the set of modes into two sets, one with $k_x>0$, the
other with $k_x<0$, and introduce $Q^\pm_{\bk\alpha}=(q_{\bk\alpha}\pm
q_{-\bk\alpha})/\sqrt{2}$, $P^\pm_{\bk\alpha}=(p_{\bk\alpha}\pm
p_{-\bk\alpha})/\sqrt{2}$. Changing the summation variable from $\bk\to-\bk$
in the $k_x<0$ part, and with $\beps_{\bk 1}=-\beps_{-\bk 1}$,
$\beps_{\bk 2}=\beps_{-\bk 2}$, $s_1=-$, $s_2=+$, Eq.(\ref{E}) can be
rewritten 
\begin{eqnarray}
\bE(\br,t)&=&\sum_{\bk,\alpha}'\sqrt{\frac{2}{\epsilon_0V}}
\beps_{\bk \alpha}\Bigg[
\left(
\omega_{k}Q_{\bk \alpha}^{s_\alpha}\cos\bk \br-P_{\bk
\alpha}^{s_{3-\alpha}}\sin\bk \br 
\right)\cos\omega_k  t
+\left(
\omega_{k}Q_{\bk \alpha}^{s_{3-\alpha}}\sin\bk \br+P_{\bk
\alpha}^{s_\alpha}\cos\bk \br 
\right)\sin\omega_k  t
\Bigg]\nonumber\,,
\end{eqnarray}
where the prime at the sum denotes summation over modes with $k_x>0$
only. Note that hereby the total number of modes is kept unchanged.
We finally perform a time dependent canonical transformation,
\begin{eqnarray}
Q_{\bk 11}(t)&=&Q_{\bk 1}^-\cos\omega_{k}t+\frac{1}{\omega_k}P_{\bk
1}^-\sin\omega_k t\\
Q_{\bk 12}(t)&=&Q_{\bk 1}^+\sin\omega_{k}t-\frac{1}{\omega_k}P_{\bk
1}^+\cos\omega_k t\,,
\end{eqnarray}
and the same set of equations, but with
$Q_{\bk 1}^\pm\to Q_{\bk 2}^\mp$, $P_{\bk 1}^\pm\to P_{\bk 2}^\mp$ for the
second polarization direction, $Q_{\bk 21}$, $Q_{\bk 22}$. This leads to
eq.(\ref{efield}).  
The advantage of this representation is that all modes
are coupled via their canonical position to the DQDs.


\begin{thebibliography}{0}
\bibitem{Bennett00} C.H. Bennett and D. P. DiVincenzo, Nature {\bf 404}, 247
(2000). 
\bibitem{Bennett93} C.H. Bennett, G. Brassard, C. Cr\'epeau, R. Jozsa,
  A. Peres, and W. K. Wootters  Phys. Rev. Lett. {\bf
70}, 1895 (1993). 
\bibitem{Jozsa02} R. Jozsa and N. Linden, quant-ph/0201143.
\bibitem{Ekert91} A. K. Ekert, Phys. Rev. Lett., {\bf 67}, 661 (1991).
\bibitem{Bouwmeester97} D. Bouwmeester, J.-W. Pan, K. Mattle, M. Eibl,
  H. Weinfurter, A. Zeilinger,  Nature {\bf 390}, 575 (1997).
\bibitem{Nielsen98} M.A. Nielsen, E. Knill, and R. Laflamme,  Nature
  {\bf 396}, 52 (1998).
\bibitem{Ursin04} R. Ursin, T. Jennewein, M. Aspelmeyer, R. Kaltenbaek,
  M. Lindenthal, P. Walther, A. Zeilinger, Nature {\bf 430}, 849 (2004). 
\bibitem{Zhao04} Z. Zhao, Y.-A. Chen, A.-N. Zhang, T. Yang, H.-J. Briegel,
  J.-W. Pan,   Nature {\bf
  430}, 54 (2004).
\bibitem{Barrett04} M.D. Barrett, J. Chiaverini, T. Schaetz, J. Britton,
  W. M. Itano, J. D. Jost, E. Knill, C. Langer, D. Leibfried, R. Ozeri,
  D. J. Wineland, Nature {\bf 429}, 737 (2004). 
\bibitem{Riebe04} M. Riebe, H. H\"affner, C. F. Roos, W. H\"ansel,
  J. Benhelm, G. P. T. Lancaster, T. W. K\"orber, C. Becher,
  F. Schmidt-Kaler, D. F. V. James, R. Blatt, Nature {\bf 429}, 734 (2004).
  \bibitem{Vandersypen01} L.M.K. Vandersypen, M. Steffen, G. Breyta,
    C. S. Yannoni, M. H. Sherwood, I. L. Chuang,  Nature {\bf 414}, 883
    (2001).  
\bibitem{Braun02} D. Braun, Phys. Rev. Lett. {\bf 89}, 277901-1 (2002).
\bibitem{Benatti03} F. Benatti, R. Floreanini, and M. Piani,
  Phys. Rev. Lett. {\bf 91}, 070402-1 (2003).  
\bibitem{Jacobczyk02} L. Jacobczyk, J. Phys. A: Math.Gen. {\bf 35}, 6383 (2002)
\bibitem{Cirone04} M.A. Cirone, G. Compagno, G.M. Palma, R. Passante, and
  F. Perisco, quant-ph/0407032.
\bibitem{Oh04} S. Oh, and J. Kim, Conference proceedings of the 2004 ERATO
Conference on Quantum Information Science, Tokyo (2004).
\bibitem{Chen05}   Y N Chen, C M Li, D S Chuu and T Brandes New J. Phys. 7
  172 (2005). 
\bibitem{Reznik03} B. Reznik,  Foundations of Physics {\bf 33}, 167 (2003). 
\bibitem{Reznik05} B. Reznik, A. Retzker, and J. Silman, Phys. Rev. A {\bf
  71}, 042104 (2005). 
\bibitem{Retzker04} A. Retzker, J.J. Cirac, and B. Reznik, quant-ph/0408059.
\bibitem{Bennett03} C.L. Bennett et al. Astrophys. J. {\bf 148} (2003).
\bibitem{Braun05} D. Braun, unpublished.
\bibitem{Braun01} D. Braun, F. Haake, and W.T. Strunz, Phys. Rev. Lett. {\bf
86}, 2913 (2001).  
\bibitem{Hayashi03} T. Hayashi, T. Fujisawa, H.D. Cheong, Y.H. Jeong, and
  Y. Hirayama, Phys. Rev. Lett. {\bf 91}, 226804-1 (2003).
\bibitem{Petta04} J. R. Petta, A. C. Johnson, C. M. Marcus, M. P. Hanson,
  and A. C. Gossard, Phys. Rev. Lett. {\bf 93}, 186802 (2004).
\bibitem{Vorojtsov04} S. Vorojtsov, E. R. Mucciolo, and H. U. Baranger,
cond-mat/0412190. 
\bibitem{Itakura03} T. Itakura and Y. Tokura, Phys. Rev. B {\bf 67},
195320 (2003). 
\bibitem{Wootters98} W.K. Wootters, Phys. Rev. Lett. {\bf 80}, 2245 (1998).
\bibitem{Mahan90} G.D. Mahan, {\em Many Particle Physics} (Plenum Press, New
  York, 1990).
\bibitem{Zurek82} W.H. Zurek, Phys. Rev. D {\bf 26}, 1862 (1982).
\bibitem{Guo97} L.-M. Duan and G.-C. Guo, quant-ph/9701020.
\bibitem{Lidar98} D.A. Lidar, I.L. Chuang, and K.B. Whaley,
Phys. Rev. Lett. {\bf 81}, 2594 (1998).
\bibitem{Braun98} D.~Braun, P.~A.~Braun,
and F.~Haake, Opt.~Comm. {\bf
179}, 411, (2000).  
\bibitem{Scully97} M.O. Scully and M.S. Zubairy, {\em Quantum Optics},
Cambridge University Press, Cambridge, UK (1997).
\end{thebibliography}
\end{document}